# Einstein, Schwarzschild, the Perihelion Motion of Mercury and the Rotating Disk Story

Galina Weinstein

Tel Aviv University

November 25, 2014

On November 18, 1915 Einstein reported to the Prussian Academy that the perihelion motion of Mercury is explained by his new General Theory of Relativity: Einstein found approximate solutions to his November 11, 1915 field equations. Einstein's field equations cannot be solved in the general case, but can be solved in particular situations. The first to offer such an exact solution was Karl Schwarzschild. Schwarzschild found one line element, which satisfied the conditions imposed by Einstein on the gravitational field of the sun, as well as Einstein's field equations from the November 18, 1915 paper. On December 22, 1915 Schwarzschild told Einstein that he reworked the calculation in his November 18 1915 paper of the Mercury perihelion. Subsequently Schwarzschild sent Einstein a manuscript, in which he derived his exact solution of Einstein's field equations. On January 13, 1916, Einstein delivered Schwarzschild's paper before the Prussian Academy, and a month later the paper was published. In March 1916 Einstein submitted to the *Annalen der Physik* a review article on the general theory of relativity. The paper was published two months later, in May 1916. The 1916 review article was written after Schwarzschild had found the complete exact solution to Einstein's November 18, 1915 field equations. Einstein preferred in his 1916 paper to write his November 18, 1915 approximate solution upon Schwarzschild exact solution (and coordinate singularity therein).

With the arrival of the centenary of the General theory of Relativity, the world finds itself celebrating a special event. Sometime in October 1915 Einstein dropped the Einstein-Grossman theory. During October 1915 Einstein adopted the determinant, $\sqrt{-g} = 1$ as a postulate, and this led him to general covariance. Starting on November 4, 1915, Einstein gradually expanded the range of the covariance of his field equations.[1] Between November 4 and November 11, 1915, Einstein realized that he did not need this postulate and he adopted it as a coordinate condition to simplify the field equations. Einstein was able to write the field equations of gravitation in a general covariant form. In the November 11, 1915 field equations the trace of the energy-momentum tensor vanishes.[2]

On November 18, 1915, Einstein presented to the Prussian Academy his paper, "Explanation of the Perihelion Motion of Mercury from the General Theory of Relativity". Einstein reported in this talk that the perihelion motion of Mercury is explained by his theory.[3] In this paper, Einstein tried to find approximate solutions to his November 11, 1915 field equations. He intended to obtain a solution, without considering the question whether or not the solution was the only possible unique solution. Astronomers found for the planet Mercury an advance of the perihelion of approximately 45" per century. If Einstein arrived at this result for the advance of the perihelion of Mercury, then his method of using an approximate rather than an exact and unique solution could not be criticized.[4]

Solving the field equations give the components of the metric tensor $g_{\mu\nu}$. In 1913 Einstein and Michele Besso tried to solve the Einstein-Grossmann "Entwurf" field equations in order to find solutions to the problem of the advance of the perihelion of Mercury in the field of a static sun. The "Entwurf" theory predicted a perihelion advance of about 18" per century instead of 45" per century. In 1915 Einstein proceeded to obtain the $g_{\mu\nu}$ for the mass of the sun by the basic method from 1913: he transferred this method to his November 18, 1915 paper and corrected it according to his new 1915 General Theory of Relativity.[5]

The solar system may be looked upon as an isolated mass, which is far away from other masses in the universe. More than 90% of the total mass of the solar system is concentrated in the sun. One can treat the planets, the masses of which are negligible as compared to the sun, as mass points moving in the *static* gravitational field of the sun. Inside the solar system one can neglect the static gravitational potential of the planets and deal only with the gravitational potential of the sun.

Einstein considered a planet, a point with negligible mass, which moves in the *static gravitational field* of a body of spherical symmetry, in a great distance from this central mass. In a very great distance from this central mass the gravitational field is so *weak* until it is not felt and we arrive back at the *Minkowski metric*. These are the conditions that Einstein imposed on the gravitational field of the sun.

The gravitational field of the sun in vacuum satisfies the following field equations (with a coordinate condition $\sqrt{-g} = 1$):[6]

$$(1) \quad \sum_{\alpha} \frac{\partial \Gamma^{\sigma}_{\mu\nu}}{\partial x_{\alpha}} + \sum_{\alpha\beta} \Gamma^{\sigma}_{\mu\beta} \Gamma^{\beta}_{\nu\alpha} = 0,$$

The left-hand side of the equation is the Ricci tensor, and it includes the metric tensor and its derivatives. Equations (1) are non-linear because of $\Gamma^{\sigma}_{\mu\nu}$.

$\Gamma^{\sigma}_{\mu\nu}$ are defined by the components of the gravitational field:[7]

$$(2) \quad \Gamma^{\sigma}_{\mu\nu} = -\begin{Bmatrix} \mu\nu \\ \alpha \end{Bmatrix} = -\frac{1}{2} \sum_{\beta} g^{\alpha\beta} \left( \frac{\partial g_{\mu\beta}}{\partial x_{\nu}} + \frac{\partial g_{\nu\beta}}{\partial x_{\mu}} - \frac{\partial g_{\mu\nu}}{\partial x_{\alpha}} \right).$$

Einstein started from the $0^{th}$ approximation: $g_{\mu\nu}$ corresponds to the special theory of relativity, to a flat Minkowski metric:

(3) $g_{\mu\nu} = \text{diag}(-1, -1, -1, +1)$,

Einstein wrote this succinctly: [8]

**(4)** $g_{\rho\sigma} = \delta_{\rho\sigma}, g_{\rho 4} = g_{4\rho} = 0, g_{44} = 1$

Here $\rho$ and $\sigma$ signify indices, 1, 2, 3; the Kronecker delta $\delta_{\rho\sigma}$ is equal to 1 or 0 when $\rho = \sigma$ or $\rho \neq \sigma$, respectively.

The approximation given in equation (4) forms the $0^{th}$ approximation. Einstein then assumed that $g_{\mu\nu}$ differ from the values given in equation (4) by an amount that is small compared to 1. He treated this deviation as a small change of "first order". Functions of $n^{th}$ degree of this deviation were treated as quantities of the $n^{th}$ order. He used equations (1) in light of equations (4) for calculation through successive approximations of the gravitational field[9] of the sun up to quantities of $n^{th}$ order.

The metric field $g_{\mu\nu}$ (the solution) has the following four properties, which are four conditions on the gravitational field of the sun: [10]

1) The solution *is static*: all components of the solution are independent of $x_4$ (time coordinate).

2) The solution $g_{\mu\nu}$ is *spherically symmetric* about the origin of the coordinate system.

3) The equations **(4)** $g_{\rho 4} = g_{4\rho} = 0$ are valid exactly for $\rho = 1, 2, 3$.

4) At infinity the $g_{\mu\nu}$ tend to the values of *the Minkowski flat metric* of special relativity given by (4).

To first order, the equations (1) and the four above conditions are satisfied through the following transformations from:[11]

**(5)** $g'_{\rho\sigma} = \text{diag}\left(-\left[1 + \frac{\alpha}{r}\right], -1 - 1, 1 - \frac{\alpha}{r}\right).$

to the following solution: [12]

**(6)** $g_{\rho\sigma} = -\delta_{\rho\sigma} - \alpha \frac{x_\rho x_\sigma}{r^3}, g_{44} = 1 - \frac{\alpha}{r}.$

The $g_{\rho\sigma}$ tends to the Minkowski metric (4) according to condition 4, and the $g_{4\rho}$ and $g_{\rho 4}$ are determined by condition 3.

The $r = \sqrt{x_1{}^2 + x_2{}^2 + x_3{}^2}$, and $\alpha = 2GM/c^2$, in Einstein's paper c = 1 is a constant determined by the mass of the static sun.

Subsequently, Einstein obtained the value for the components of the gravitational field of the static sun to the second order approximation. He wrote equations of motion for a point mass moving in the gravitational field of the sun. A planet in a free fall in the gravitational field of the sun moves on a geodesic line according to the geodesic equation: [13]

$$\text{(7)} \quad \frac{d^2 x_\nu}{ds^2} = \sum_{\sigma\tau} \Gamma^\nu_{\sigma\tau} \frac{dx_\sigma}{ds} \frac{dx_{\nu\tau}}{ds}.$$

Einstein calculated the equations of the geodesic lines and compared them with the Newtonian equations of the orbits of the planets in the solar system. Hence, he checked whether there is correspondence between general relativity and Newtonian theory.

In Newtonian theory the gravitational attraction is a central force, and all planets move in a constant plane around the sun. Hence in polar coordinates the motion of this plane is dependent on the distance $r$ of the planet from the center, and $\phi$ the angle between the line that connects the planet to the center and a line that is chosen arbitrarily. One obtains the orbit equation, and $r$ as a function of $\phi$ (the distance of the planet from the sun at any given angle). The solution of the Newtonian orbit equation is the equation of an ellipse – an orbit in the plane, and the eccentricity $e$ determines the characteristic of the elliptic orbit.

The perihelion of the orbit is the point in which the planet is closest to the sun. This point is found on the major axis of the ellipse, its longest diameter, the line that runs through the centre and both its foci. This major axis was found to slowly turn around the sun; and the perihelion turned as well. This is the precession of the perihelion, and it is more pronounced the more the eccentricity $e$ is larger.

In Einstein's theory the geodesic equation leads to an orbit equation. The geodesic equation led Einstein to a relativistic equation of the orbit.[14] Einstein found that the difference between the Newtonian orbit equation and the relativistic orbit equation was in an additional term: $2GM/c^2 r^3$ that appears in the relativistic equation. He treated first the Newtonian solution to this equation as a first approximation. He then checked, what was the size of the correction that resulted from the addition of this term? He integrated the Newtonian orbit equation first. The Newtonian solution to the Newtonian orbit equation describes an ellipse of a planet, for which the direction of the major axis and the perihelion should both stay fixed.

Einstein then added the perturbation of the additional term $2GM/c^2r^3$ to this solution in order to see whether the turning of the perihelion resulted from this additional term in the relativistic equation. If this was indeed the result, then the precession of the perihelion would turn to be a result of a relativistic effect, and this was the first triumph of Einstein's 1915 theory. Einstein concluded his scheme by saying, "The calculation yields, for the planet Mercury, an advance of the perihelion of 43″ per century, while the astronomers indicated **45″ ± 5″** as the unexplained remainder between observations and the Newtonian theory. This means full compatibility".[15] A great triumph for Einstein's November 1915 theory.

Between November 18 and November 25 Einstein found that he could write the field equations with an additional term on the right hand side of the field equations involving the trace of the energy-momentum tensor, which now need not vanish. These were the final November 25, 1915 field equations.

Einstein's field equations are non-linear partial differential equations of the second rank. This complicated system of equations cannot be solved in the general case, but can be solved in particular simple situations. The first to offer an exact solution to Einstein's November 18, 1915 field equations was Karl Schwarzschild, the director of the Astrophysical Observatory in Potsdam.

On December 22, 1915 Schwarzschild wrote Einstein from the Russian front: "As you can see, the war is kindly with me, giving me fire, in spite of fierce gunfire, allowing in the very terrestrial distance, this stroll in your land of ideas". Schwarzschild set out to rework Einstein's calculation in his November 18 1915 paper of the Mercury perihelion problem.[16]

He first responded to Einstein's solution for the first order approximation, and found another first-order approximate solution. Schwarzschild told Einstein that the problem would be then physically undetermined if there were a few approximate solutions. Subsequently, Schwarzschild presented a complete solution. He said he realized that there was only one line element, which satisfied the four conditions imposed by Einstein on the gravitational field of the sun, as well as Einstein's field equations from the November 18 1915 paper. The problem with Schwarzschild's line element was that a mathematical singularity was seen to occur at the origin.

Schwarzschild considered a body, the origin of the coordinates is its geometric center. If one assumes isotropy of space and *a static solution*, then there exists spherical symmetry around the center; and one can work with a *system of spherical coordinates* R, ϑ, φ. The symmetry of the solution means that the variables are independent of the angular coordinates ϑ, φ. Since the solution is static, there is no dependence on time, and thus only R is an independent variable, the distance from the center. Schwarzschild wrote to Einstein the following:[17]

"Let,

$x_1 = r\cos\varphi\cos\vartheta, \quad x_2 = r\sin\varphi\cos\vartheta, \quad x_3 = r\sin\vartheta,$

Consider,

$R = (r^3 + \alpha^3)^{1/3} = r\left(1 + \frac{1}{3}\frac{\alpha^3}{r^3} + \cdots\right),$

then the line element becomes:

**(8)** $ds^2 = \left(1 - \frac{\gamma}{R}\right)dt^2 - \frac{dR^2}{1 - \frac{\gamma}{R}} - R^2(d\vartheta^2 + \sin^2\vartheta d\varphi^2).$"

where, $\gamma = 2GM/c^2$ ($c^2 = 1$).

Schwarzschild wrote that R, $\vartheta$, $\varphi$ are not "allowed" coordinates, with which the field equations can be formed, because these spherical coordinates do not satisfy the coordinate condition from Einstein's November 18 paper, $\sqrt{-g} = 1$. The above line element expressed itself as the best in spherical coordinates. "The equation of the orbit remains exactly as" Einstein "obtained in the first approximation",[18] but at the seemingly unacceptable cost of the choice of non-"allowed" coordinates, and the mathematical singularity that occurred in the solution when R = 0. If we consider (8), then one easily arrives at Einstein's relativistic equation of the orbit (orbit equation + $2GM/c^2r^3$), and this equation gives the observed precession of the perihelion of Mercury.

Einstein replied to Schwarzschild on December 29, 1915 and told him that his calculation proving uniqueness proof for the problem is very interesting. "I hope you publish the idea soon! I would not have thought that the strict treatment of the point-problem was so simple".[19] Subsequently Schwarzschild sent Einstein a manuscript, in which he derived his solution of Einstein's November 18, 1915 field equations for the field of a single mass.[20]

Schwarzschild sitting in the Russian front, found a "simple trick" that allowed him to avoid the problem of the non-"allowed" coordinates: *The new variables are therefore spherical coordinates of the determinant 1*".[21] Einstein's field equations (1) and the coordinate condition, $\sqrt{-g} = 1$ from his November 18 paper were satisfied. But returning back to the "standard" (that is, non-"allowed") spherical coordinates, we arrive at the exact solution to Einstein's problem, and to the mathematical singularity in the solution when R = 0.

Einstein received the manuscript by the beginning of January 1916, and he examined it "with great interest". He told Schwarzschild that he "did not expect that one could formulate so easily the rigorous solution to the problem". On January 13, 1916,

Einstein delivered Schwarzschild's paper before the Prussian Academy with a few words of explanation.[22] Schwarzschild's paper, "On the Gravitational Field of a Point-Mass according to Einstein's Theory" was published a month later.[23]

In March 1916 Einstein submitted to the *Annalen der Physik* a review article on the general theory of relativity, "The Foundation of the General Theory of Relativity". The paper was published two months later, in May 1916. The 1916 review article was written *after* Schwarzschild had found the complete exact solution (8) to Einstein's November 18, 1915 field equations. Even so, in his 1916 paper, Einstein preferred not to base himself on Schwarzschild's exact solution (which did not satisfy the coordinate condition $\sqrt{-g} = 1$), and he returned to his first order approximate solution (6) from his November 18, 1915 paper.

In his 1916 review paper Einstein chose the coordinates such that $\sqrt{-g} = 1$. He explained: "I will therefore give below all relations in the simplified form, which this specialization of the choice of coordinates brings with it. It will then be easy to access to the *generally* covariant equations, if this seems desirable in a special case".[24] Einstein then explained why this choice does not mean a "partial abandonment of the general postulate of relativity". The reason is that we do not ask: what are the laws of nature which are covariant for transformations of determinant 1? We rather first ask: "What are the generally covariant laws of nature?"[25] Only after formulating these, we then simplify their expression by a special choice of *a reference system* $\sqrt{-g} = 1$. Hence, following the stages from the beginning of November 1915, Einstein adopted the determinant in equation $\sqrt{-g} = 1$ as a *postulate*. Then in the November 11 paper he adopted it as a *coordinate condition*, and in the 1916 review article he expressed his field equations with respect to the *special reference system* $\sqrt{-g} = 1$.

In section §3 of his 1916 paper Einstein presented the rotating disk thought experiment.[26] And in doing so, Einstein used a coordinate-dependent description of the kind one finds in his special relativity papers and in his 1911-1912 gravitation papers.[27] But the initial motivation for presenting the rotating disk thought experiment in 1916 was to show that coordinates of space and time have no direct physical meaning; since coordinates have no direct physical meaning Euclidean Geometry breaks down. In the final part of the 1916 paper Einstein came back to the rotating disk problem with which he opened his 1916 paper. What happens to the length of a rod and the measurement of time in the presence of a gravitational field? In order to answer this question Einstein analyzed the disk problem using the metric equation and the first order approximate solution from his November 18, 1915 paper.

In section §3 of his 1916 paper Einstein considered two systems of reference, the Galilean K and the one K', which is in uniform rotation relative to K. The origin of both systems, as well as their axes of Z, permanently coincide one with another. The circle of a disk around the origin in the X, Y plane of K is regarded at the same time as a circle of a disk in the X', Y' plane of K'. We imagine that the circumference and diameter of this circle are measured with a unit rod (infinitely small relative to the radius), and we form the quotient of the two results. If the experiment is performed with a measuring rod at rest relative to K, the quotient will be $\pi$. With a measuring rod at rest relative to K', the quotient will be greater than $\pi$. If the whole process of measurement is viewed from K, the periphery undergoes a Lorentz contraction, while the measuring rod applied to the radius does not. It follows that therefore the lengths measurements have no direct meaning and Euclidean geometry does not apply to K'.

After propounding on lengths measurements of the circle of the disk, Einstein discusses time measurements. Einstein imagined two clocks of identical constitution placed, one at the origin of coordinates, and the other at the periphery of the circle of the disk. Both clocks are observed from K. According to time dilation, judged from K, the clock at the periphery of the circle of the disk goes more slowly than the other clock at the origin, because the clock at the circumference is in motion and the one at the origin is at rest. An observer who is located at the origin, and who is capable of observing the clock at the circumference by means of light, would be able to see the periphery clock lagging behind the clock beside him. He will interpret this observation as showing that the clock at the periphery goes more slowly than the clock at the origin. He will thus define time in such a way that the rate of the clock depends upon its location.

Hence, when we measure the circumference of the circle of the disk of K' from the system K, then the measuring-rod applied to the periphery undergoes a Lorentzian contraction, while the one applied along the radius does not; and when we require measurement of time events in K', then judged from K, the clock at the periphery of the circle of the disk goes more slowly than the other clock at the origin. Euclidean geometry breaks down in the system K', and so too we are unable to introduce a time corresponding to physical requirements in K', indicated by clocks at rest relatively to K.

The story Einstein tells in section §3 of his 1916 paper, is based on an old manner of expression, a coordinate-dependent description (the equivalence principle): "Can an observer at rest relatively to K' infer that he is on a 'really' accelerated reference system? The answer to this question is negative; because the above-mentioned behavior of the freely moving masses relative to K' can be equally interpreted in the following way. The reference system K' is unaccelarated; but in the considered space-time regions there is a gravitational field, which generates the accelerated motion of the bodies with respect to K'." [28]

Fairly soon afterwards, Einstein explained the reason for presenting the disk story: "In the general theory of relativity, space and time cannot be defined in such a way that spatial coordinate differences be directly measured by the unit measuring rod, and time by a standard clock".[29]

Einstein returned to the rotating disk thought experiment towards the end of his 1916 paper. But he now started with the metrical properties of space-time and used the first order approximate solution (6) from his November 18, 1915 paper in order to demonstrate that a gravitational field *changes spatial dimensions and the clock period*.

Consider the line element:

**(9)** $ds^2 = g_{\mu\nu} dx_\mu dx_\nu$

and a unit-measuring rod laid "parallel" to the x-axis.

Then, $ds^2 = -1$; $dx_2 = dx_3 = dx_4 = 0$. Therefore, equation (9) gives, $-1 = g_{11} dx_1^2$.

Suppose the unit-measuring rod laid "parallel" to the x-axis also lies *on* the x-axis. In this case, the first of equations (6) gives: [30]

**(10)** $g_{11} = -\left(1 + \dfrac{\alpha}{r}\right).$

Equation (10) and equation (9) in the form: $-1 = g_{11} dx_1^2$, yield,

**(11)** $dx = 1 - \dfrac{\alpha}{2r}.$

Einstein readily derived equation (11) using (9) and (6), and concluded: "The unit measuring rod therefore appears a little shortened with respect to the coordinate system by the presence of the gravitational field, if it is laid in the radial direction".

As to the length of a measuring rod in the tangential direction: we set $ds^2 = -1$; $dx_1 = dx_3 = dx_4 = 0$; $x_2 = r$, $x_1 = x_3 = 0$. Therefore, equation (8) gives,

(12) $-1 = g_{22} dx_2^2 = -dx_2^2$.

With the tangential position, therefore, the gravitational field of the point mass has no influence on the length of a rod.

Let us rewrite equation (11) with the gravitational potential Φ in the following form:

**(13)** $dx = 1 + \dfrac{\Phi}{c^2}.$

In the 1916 paper, starting from the line element (9), Einstein derived gravitational redshift. Einstein considered the rate of a unit clock, which is arranged at rest in a static gravitational field. For the clock period we set, $ds = 1$; $dx_1 = dx_2 = dx_3 = 0$. Thus, $1 = g_{44} dx_4^2$. Consider,

$$(14)\ dx_4 = \frac{1}{\sqrt{g_{44}}} = \frac{1}{\sqrt{1 + (g_{44} - 1)}} = 1 - \frac{g_{44} - 1}{2}.$$

Einstein concluded, "The clock goes then more slowly if it is placed near ponderable masses. It follows that the spectral lines of light reaching us from the surface of large stars must appear displaced towards the red end of the spectrum".[31] This is *gravitational redshift of light*.

Let $d\tau_0$ be a volume element in the local reference system where special relativity applies, then,[32]

$$(15)\ d\tau_0 = \sqrt{-g}\, d\tau.$$

If $\sqrt{-g} = 1$, then $d\tau_0 = d\tau$.

Let us return to Einstein's definition, the equivalence principle, applicable to local systems: Experiments in a sufficiently small free falling system, over a sufficiently short time interval, give results that are indistinguishable from those of the same experiments done in an inertial frame in which special relativity applies.

Consider the following Gedanken-experiment. Imagine two systems. One system is Einstein's 1907 imaginary man falling from the roof in a gravitational field. In the other system there is a man at rest in a gravitational field. Consider the local inertial system of special relativity: Imagine the man at the moment he starts to fall from the roof. At this infinitesimally initial moment, he is still at rest. Both men at both systems are thus at rest at this very moment. The worldlines of these men are comprised of the time intervals according to (15):

$$(16)\ \sqrt{g_{44}'}\, d\tau' = \sqrt{g_{44}}\, d\tau.$$

In the local inertial system all the diagonal components of the metric tensor are constants and according to (3) they are equal to 1, and the off diagonal components are zero. And thus $g_{44} = 1$, and we arrive at an equation (15).

Einstein wrote in 1911: If we measure time in a lower gravitational potential with a clock $U$, we must measure the time in a higher gravitational potential with a clock that goes $1 + \Phi/c^2$ slower than the clock U if you compare it with the clock U in the same place.[33]

In the limit of weak gravitational fields, Einstein expected that $g_{44}$ would tend to the above factor. Thus:

**(17)** $\sqrt{g_{44}} = \sqrt{1 + \frac{\Phi}{c^2}}$.

Further, Einstein derived *bending of light*, the deflection of a ray of light passing by the sun with Huygens principle and equation (6). [34] Already in the November 18, 1915 paper Einstein had found that his theory laid so far, could lead to another result that occupied him since 1907; it could produce, "a somewhat different influence of the gravitational field on the light ray as in my earlier work, because the velocity of light is determined by the equation

**(18)** $[ds^2 =] \sum g_{\mu\nu} dx_\mu dx_\nu = 0$.

By application of the Huygens principle, we find from equations (18) and (6) through a simple calculation, that a light ray passing at a distance Δ undergoes an angular deflection […]". [35]

Einstein ended his paper with the final equation from his November 18 paper, the equation for *the perihelion advance of Mercury* in the sense of motion after a complete orbit. [36] And he only mentioned in a footnote, [37] "With respect to the calculation, I refer to the original treatises": Einstein's November 18 paper and Schwarzschild's 1916 paper.

---

[1] Einstein, Albert (1915a). "Zur allgemeinen Relativitätstheorie." *Königlich Preußische, Akademie der Wissenschaften* (Berlin). *Sitzungsberichte*, 778-786.

[2] Einstein, Albert (1915b). "Zur allgemeinen Relativitätstheorie. (Nachtrag)." *Königlich Preußische Akademie der Wissenschaften* (Berlin). *Sitzungsberichte*, 799-801.

[3] Einstein, Albert (1915c). "Erklärung der Perihelbewegung des Merkur aus der allgemeinen Relativitätstheorie." *Königlich Preußische Akademie der Wissenschaften* (Berlin). *Sitzungsberichte*, 831-839.

[4] Earman, John and Janssen, Michel (1993). "Einstein's Explanation of the Motion of Mercury's Perihelion." In John Earman, Michel Janssen, John D. Norton (ed), *The Attraction of Gravitation: New Studies in the History of General Relativity, Einstein Studies*, MA: Springer, 129-172; 141.

[5] *The Collected Papers of Albert Einstein. Vol. 4: The Swiss Years: Writings, 1912–1914* (*CPAE* 4). Klein, Martin J., Kox, A.J., Renn, Jürgen, and Schulmann, Robert (eds.), Princeton: Princeton University Press, 1995, "The Einstein-Besso Manuscript on the Motion of the Perihelion of Mercury", 349-351.

[6] Einstein 1915c, 832.

[7] Einstein 1915a, 783.

[8] Einstein 1915c, 832.

[9] Earman and Janssen say that it is not clear whether Einstein meant here the metric field of the sun or the components of the gravitational field of the sun. Earman and Janssen 1993, 142.

[10] Einstein 1915c, 833.

[11] Earman and Janssen 1993, 143-144.

[12] Einstein 1915c, 833.

[13] Einstein 1915c, 835.

[14] Einstein 1915c, 837.

[15] Einstein 1915c, 839.

[16] Schwarzschild to Einstein, 22 December 1915, *the Collected Papers of Albert Einstein. Vol. 8: The Berlin Years: Correspondence, 1914–1918* (*CPAE* 8), Schulmann, Robert, Kox, A.J., Janssen, Michel, Illy, Jószef (eds.), Princeton: Princeton University Press, 2002, Doc. 169, note 5.

[17] Schwarzschild to Einstein, 22 December 1915, *CPAE* 8, Doc. 169.

[18] Schwarzschild to Einstein, 22 December 1915, *CPAE* 8, Doc. 169.

[19] Einstein to Schwarzschild, 29 December 1915, *CPAE* 8, Doc. 176.

[20] *CPAE* 8, note 1, 242.

[21] Schwarzschild, Karl (1916), "Über das Gravitationsfeld eines Massenpunktes nach der Einsteinschen Theorie." *Königlich Preußische Akademie der Wissenschaften (Berlin). Sitzungsberichte*, 189-196;191.

[22] Einstein to Schwarzschild, 9 January 1916, *CPAE* 8, Doc. 181.

[23] Schwarzschild 1916.

[24] Einstein, Albert (1916). "Die Grundlage der allgemeinen Relativitätstheorie." *Annalen der Physik* 49, 769-822; 801.

[25] Einstein, 1916, 789.

[26] Einstein, 1916, 774-775.

[27] Einstein, Albert (1911). "Uber den Einfluβ der Schwerkraft auf die Ausbreitung des Lichtes.", *Annalen der Physik* 35, 898-908; (1912a). "Lichtgeschwindigkeit und Statik

des Gravitationsfeldes." *Annalen der Physik* 38, 355-369; (1912c), "Zur Theorie des statischen Gravitationsfeldes." *Annalen der Physik* 38, 1912, 443-458.

[28] Einstein 1916, 772-773.

[29] Einstein 1916, 775.

[30] Einstein 1916, 819.

[31] Einstein 1916, 820.

[32] Einstein 1916, 789.

[33] Einstein 1911, 906.

[34] Einstein 1916, 821.

[35] Einstein 1915c, 834.

[36] Einstein 1916, 822, 1915c, 839.

[37] Einstein 1916, 822.